# Transport and spin conversion of multi-carriers in semimetal bismuth

Hiroyuki Emoto [1,2], Yuichiro Ando [1], Gaku Eguchi [1,5], Ryo Ohshima [1,2]
Eiji Shikoh [3], Yuki Fuseya [4], Teruya Shinjo [1], Masashi Shiraishi [1,#]

1. Department of Electronic Science and Engineering, Kyoto University, Nishikyo-ku, Kyoto, 615-8510, Japan
2. Graduate School of Engineering Science, Osaka University, Toyonaka, Osaka 560-8530, Japan.
3. Graduate School of Engineering, Osaka City University, Sumiyoshi-ku, Osaka, 558-8585, Japan.
4. Department of Engineering Science, University of Electro-Communications, Chofu, Tokyo 182-8585, Japan
5. Institute of Solid State Physics, TU-Wien, 1040 Vienna, Austria

# Corresponding author: Masashi Shiraishi (mshiraishi@kuee.kyoto-u.ac.jp)

## Abstract

In this paper, we report on the investigation of (1) the transport properties of multi-carriers in semi-metal Bi and (2) the spin conversion physics in this semimetal system in a ferrimagnetic insulator, yttrium-iron-garnet. Hall measurements reveal that electrons and holes co-exist in the Bi, with electrons being the dominant carrier. The results of a spin conversion experiment corroborate the results of the Hall measurement; in addition, the inverse spin Hall effect governs the spin conversion in the semimetal/insulator system. This study provides further insights into spin conversion physics in semimetal systems.

# I. Introduction

Semimetal bismuth (Bi) is one of the most intensively studied elements in solid-state physics, and a wide variety of attractive physics, such as the Nernst-Ettingshausen effect [1], the Shbnikov-de Haas oscillation [2], the de Haas-van Alphen effect [3], the Seebeck effect [4] and so on, has been discovered by using Bi. Studies of Bi in solid-state physics [5] have demonstrated that the low-energy effective Hamiltonian of Bi is given by the Dirac Hamiltonian with a small band gap (approximately 10 meV) at the L-point [6-8], strong diamagnetism [9] and a large spin-orbit interaction (SOI, approximately 1.8 eV) [10]. Among these characteristics, the correlation between a possible long momentum relaxation length due to the Dirac-like linear band structure and the quite large SOI resulting in rapid spin relaxation, attracts strong attention in spintronics because these properties govern spin transport and spin conversion in Bi.

Electric spin conversion is the conversion from an electric current to a spin current and *vice versa*, such as the spin Hall effect and the inverse spin Hall effect (ISHE). Here, the spin carrier species, i.e., electrons or holes, govern the sign of the electromotive forces generated by the ISHE. Charge transport properties in a semimetal system such as Bi, where electrons and holes co-exist, can be understood from the longitudinal and transverse resistivity, and a recent relevant analysis enables obtaining further insight into the multi-carrier transport [11]. Hence, a combination of electric spin conversion and resistivity measurements enables an understanding of the physics involving a correlation between spin and charge transport in semimetals. An example of the experimental demonstration of spin conversion in Bi is the ISHE [12,13], and these studies are motivated by the large SOI of Bi. Another notable study is the inverse Rashba-Edelstein effect (IREE) [14,15] appearing at a Bi/Ag interface, which has been attributed to a large Rashba splitting at the interface [16]. However, further discussion of the spin conversion physics that considers multi-carrier transport in Bi remains lacking.

The purpose of this study is to investigate spin and charge transport and their conversion in Bi via the Hall and the ISHE measurements, where the former enables investigation of charge transport and the latter enables the study of spin transport and its conversion properties. We chose a Bi/yttrium-iron-garnet (YIG) heterostructure, in which Bi on the YIG was polycrystalline. The introduction of the magnetic insulator, YIG instead of a conductive ferromagnetic metal such as NiFe, enables to bypass the problem of the superposition of unwanted electromotive forces under FMR from the conductive ferromagnet on the spin conversion signals [17,18]. In addition, the semimetal/insulator interface can induce the Rashba field, which enables detection of the IREE.

## II. Experiments

Bi thin films were fabricated on single crystalline YIG, where the thickness of the Bi was varied from 5 to 60 nm. Figure 1(a) shows an X-ray diffraction $\theta$-$2\theta$ pattern of the Bi sample (20 nm in thick), where Cu-K$\alpha$ radiation was used. The peaks labeled (003), (006) and (009) were observed, whereas the peaks labeled (012) and (110) were not. This directly implies that (001) is the preferential orientation. By contrast, the appearance of the (104) peak indicates that the Bi is not perfectly aligned to the (001) direction. The result tells us that (1) 80% of the Bi is polycrystalline and that (2) the Bi is rhombohedral and oriented along (001) with 20% of the (104) oriented crystal. Electric transport properties of the Bi/YIG film were determined implemented using the conventional four-probe technique with a commercial apparatus

(Quantum Design PPMS). Electrical contacts were made using room-temperature-cured silver paste, and the process was performed under atmospheric conditions. The Hall measurement was implemented in the temperature range of 1.8 to 300 K to investigate charge carrier transport properties, and the thickness of the Bi varied from 8 to 70 nm. We introduced spin pumping for injecting a pure spin current from the YIG to the Bi to study the spin-to-charge conversion physics in this semimetal. Spin pumping is a method of spin injection from a ferromagnet/ferrimagnet to a nonmagnet under ferromagnetic resonance (FMR) of the ferromagnet. Under FMR, spin angular momenta in the ferromagnet are transferred to the conducting electrons in the nonmagnet via the $s$-$d$ coupling, resulting in spin accumulation and generation of a pure spin current in the nonmagnet [19]. In the spin pumping measurements, the Bi/YIG sample was placed in a nodal position of a $TE_{011}$ cavity of an electron spin resonance (ESR) system (JEOL FA-200), where the alternating electric and magnetic field components were a minimum and a maximum, respectively (the microwave frequency was set to be 9.12 GHz). The excitation power of the microwave was applied up to 1 mW. An external static magnetic field for obtaining the FMR was applied at an angle, $\theta_H$, as shown in Fig. 1(b). All measurements of the spin pumping were performed at room temperature (RT).

In order to distinguish the ISHE and the IREE in the Bi/YIG, the Bi thickness dependence of a generated electric current by these effects provides important information. As illustrated in Fig. 1(c), the physical origins of the ISHE and the IREE are essentially different. In the ISHE, spin-dependent scattering gives rise to conversion of a spin current to a charge current, and the length scale of the conversion is determined by the spin diffusion length. Hence, the thickness dependence of the generated charge current exhibits a characteristic behavior as shown in Fig.

1(c). However, the IREE is ascribed to the interfacial Rashba effect of Bi, resulting in a constant amount of a charge current generated via the spin conversion. Thus, the total generated charge current does not exhibit a dependence on the thickness of the Bi.

**III. Results and Discussion**

Concerning the electric transport properties, a sharp increase in the sheet resistivity ($R_{xx}$) was observed with decreasing Bi thickness (see Fig. 2(a)). Decrease in the sheet Hall coefficient ($R_H = R_{xy}/B$, $B = \mu_0 H$) and in the Hall mobility ($\mu_H = R_H/R_{xx}$) were also observed below 20 nm. These results indicate that the increase in $R_{xx}$ is attributable to the decrease in $\mu_H$, implying a significant decrease in total mobility. The absence of magnetoresistance (MR) for ≤ 8 nm, as shown in Fig. 2(b), also supports the behavior because the amplitude of the MR is scaled by $\mu_H B$. The MR at 70 nm and 20 nm was analyzed using the two-carrier model of electrons and holes:

$$\frac{R_{xx}(B) - R_{xx}(0)}{R_{xx}(B)} = \frac{(N^2 - 1)M^2(1 - M^2)\mu_H^2 B^2}{(2M + N + NM^2)^2 + (N^2 - M^2)(1 - M^2)\mu_H^2 B^2}, \quad (1)$$

where, $n_1$ and $n_2$ are the sheet carrier densities ($n_1 > n_2$), $N = (n_1 - n_2)/(n_1 + n_2)$, $\mu_1$ and $\mu_2$ are the mobilities, $M = (\mu_1 - \mu_2)/(\mu_1 + \mu_2)$, and

$$\mu_H = \frac{\mu_1 + \mu_2}{2} \cdot \beta = \frac{\mu_1 + \mu_2}{2} \cdot \frac{2M + N + NM^2}{N + M}. \quad (2)$$

Note that the values of $\mu_H$ is the experimental value, presented in the Fig. 2(a), where $\mu_H$ is $0.02302 \pm 0.00001$ m²/Vs for 70 nm, and $0.01273 \pm 0.00001$ m²/Vs for 20 nm. Furthermore, note that the mobility is defined to be negative for electron and positive for holes. The MR was thus determined based on the two active parameters, namely, $N$ and $M$, resulting in $N=0.7848 \pm$

0.0003 and $M=-1.7250\pm0.0008$ for 70 nm, and $N=0.9583\pm0.0001$ and $M=-1.1992\pm0.0002$ for 20 nm. Within the model, the sheet Hall coefficient, $R_H$, is expressed as

$$R_H = R_0 \cdot \alpha = R_0 \frac{2M + N + NM^2}{(N+M)^2}, \quad (3)$$

where $R_0 = [(n_1 + n_2)q_1]^{-1}$ and $q_1 = \pm e$ is the charge of the majority carrier ($n_1$). Here, the positive or negative charge is determined from the relation

$$\frac{\mu_1 - \mu_2}{2} = \frac{\mu_1 + \mu_2}{2} \cdot M, \quad (4)$$

because the value becomes negative for $q_1=-e$ and positive for $q_1=+e$. For the present case, the values were -0.1132 for 70 nm and -0.0593 for 20 nm; thus $q_1=-e$. The majority carrier was an electron, which is the central result of the Hall measurement. The contour plot of the Hall factor $\alpha$ and what was determined from $N$ and $M$ are shown in Fig. 2(c). $\alpha$ exhibited negative values and was consistent with the observed positive $R_H$ values. Finally, from the experimentally determined parameters $R_H$, $\mu_H$, $N$, and $M$, the two-carrier-transport-related values were determined as $n_1 = [(1.43 \pm 0.02) \times 10^{18}]/m^2$, $n_2 = [(0.18 \pm 0.02) \times 10^{18}]/m^2$, $\mu_1 = [-(0.047 \pm 0.002)]m^2/Vs$, and $\mu_2 = [(0.179 \pm 0.002)]m^2/Vs$ for 70 nm, and $n_1 = [(1.76 \pm 0.03) \times 10^{18}]/m^2$, $n_2 = [(0.04 \pm -0.03) \times 10^{18}]/m^2$, $\mu_1 = [-(0.0099 \pm 0.0010)]m^2/Vs$, and $\mu_2 = [(0.1088 \pm 0.0010)]m^2/Vs$ for 20 nm. Further details of the transport measurements are provided in the *Supplemental Materials* [20]

To understand the spin conversion properties at the Bi/YIG interface, spin pumping experiments were implemented. FMR signals from the YIG at $\theta_H=0$ and 180 degrees were clearly found when the resonant field was 245 mT (see Fig. 3(a)). Electromotive forces from the Bi under the FMR were observed at $\theta_H=0$ and 180 degrees, where the polarity of the signal was reversed (see Fig. 3(b)). Because a thermally induced signal can be superimposed in the signals, we took an average of both signals (the black solid line in Fig. 3(c)). The power dependence of the averaged electromotive forces is shown in Figs. 3(c) and (d), and a linear dependence was observed. These findings are in good accordance with the physical features expected in the ISHE and the IREE. The polarity of the electromotive forces due to the spin conversion indicates that the spin carrier is an electron in all cases, which is consistent with the result of the Hall measurement. In other words, it is not the four-fold difference in carrier mobilities but the ten-fold difference in the carrier densities that governs the spin conversion.

To reveal the physical origin of the observed signals, the thickness dependence of the electromotive forces from the Bi was investigated. Note that the spin conversion mechanisms of the ISHE and the IREE are essentially different. Because the interfacial Rashba SOI generates the IREE, a charge current converted from a pumped spin current does not depend on the Bi thickness. By contrast, the bulk SOI of Bi induces the ISHE; thus the charge current increases along with the Bi thickness as mentioned in Section II. The thickness dependence of the charge current (from 10 to 60 nm) is presented in Fig. 4. As shown in the figure, the charge current monotonically increased as a function of the Bi thickness, which is apparently different from the behavior expected from the IREE. The power dependence and the angular dependence of the electromotive forces are certainly consistent with the characteristics of the ISHE. Therefore, we conclude that the spin conversion in Bi/YIG, where the electromotive forces are not generated from a spin source, is attributed to the ISHE. The charge current as a function of the Bi thickness is described as

$$I_c = w\theta_{SHA}(\frac{2e}{\hbar})\lambda_s \tanh(\frac{d}{2\lambda_s})J_s^0,\qquad(1)$$

where $w$ is the width of the Bi thin film (1.5 mm), $d$ is the thickness of the Bi, $e$ is the electric charge, $\hbar$ is the Dirac constant, $\lambda_s$ is the spin diffusion length of the Bi, $J_s^0$ is the spin current density (estimated to be $2.89 \times 10^{-10}$ J/m$^2$ in this study), and $\theta_{SHA}$ is the conversion efficiency of a spin current to a charge current, i.e., the spin Hall angle [26]. We performed a theoretical fitting of the experimental data by setting $\lambda_s$ as a free parameter. Notable is that the spin diffusion length to the different thickness samples is assumed to be identical because of the following reason: the spin diffusion length is independent on the momentum relaxation time under the D'yakonov-Perel type spin relaxation mechanism. The Bi in this study is polycrystalline, which allows the D'yakonov-Perel mechanism due to its heavy atomic mass. The effective mass of electron can be the same as a function of the thickness of Bi, hence the spin diffusion length is identical. As shown in Fig. 4, the magnitude of the charge current, $I_c$, is nicely fitted by Eq. (1) resulting in a spin diffusion length to be approximately 20 nm at RT. In the previous study, the spin diffusion length of Bi on NiFe was estimated to be 8 nm [13]; however the present data cannot be explained if the spin diffusion length is 8 nm (see Fig. 4). The discrepancy is attributable to the difference in the film qualities, i.e., polycrystalline in this study and amorphous in the previous study. In the theoretical fitting, the spin Hall angle was estimated to be 0.00012, and is much smaller than that of Pt (0.1 [21]), although the spin-orbit interaction of Bi is twice as large as that of Pt. In previous studies, the spin Hall angle of Bi was comparatively large (0.02 [13]), and the Bi was amorphous-like and the resistivity was one order of magnitude larger than that in the Bi of this study. The disorder in the amorphous Bi easily induced the Elliot-Yafet spin relaxation. The ISHE is a spin-dependent scattering phenomenon due to momentum scattering centers; i.e., momentum relaxation induces the ISHE. Hence, the transporting spins in the amorphous Bi are conserved by momentum relaxation through a charge current by the ISHE before they lose their spin coherence via the SOI. However, as mentioned above, the Bi in this study was polycrystalline and the D'yakonov-Perel type SOI can govern the spin relaxation because of the heavy mass of Bi. In addition, transporting spin angular momenta in the Bi are lost as a result of the D'yakonov-Perel type SOI (not the momentum relaxation). Consequently, the total amount of spins converted to a charge current by the ISHE in the Bi of this study is much smaller than that in the amorphous Bi, resulting in a decrease in the total amount of a charge generated by the ISHE and the small spin Hall angle in the Bi.

### IV. Summary

In summary, we investigated the transport properties of multi-carriers in thin-film semimetallic Bi that strongly govern spin conversion in the film. As measured, the electron is the dominant carrier in the Bi at RT, which is consistent with the polarity of the EMF from the Bi observed in the spin conversion experiments. The Bi thickness dependence of the generated electric current by the spin pumping and spin conversion in the Bi tells us that the spin conversion physics is governed by the ISHE, not by the IREE. This study provides significant materials for understanding further insight into the spin conversion physics of semimetal systems.

**Acknowledgements**

This work was partly supported by a Grant-in-Aid for Scientific Research on Innovative Areas, "Nano Spin Conversion Science" (grant No. 26103003) and by a Grant-in-Aid for Scientific Research (A) (grant No. 15H02108).

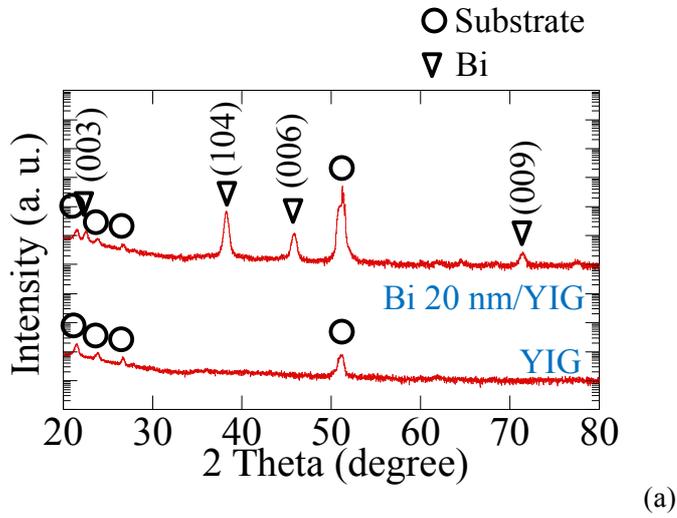

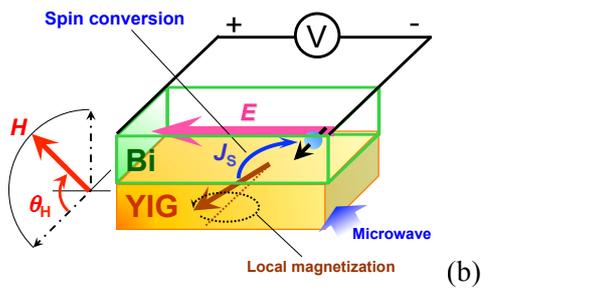

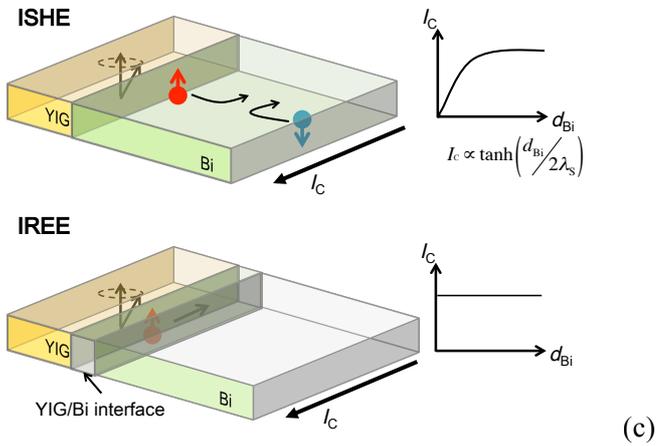

**Figure 1** (a) X-ray diffraction patterns of the Bi(20 nm)/YIG and of the YIG. The crystalline peaks are identified as shown in the figure. (b) Schematic image of the spin conversion device. An external magnetic field, H, is applied as shown in the figure for the FMR of the YIG. (c) Illustration of different spin conversion principles in the ISHE and the IREE.

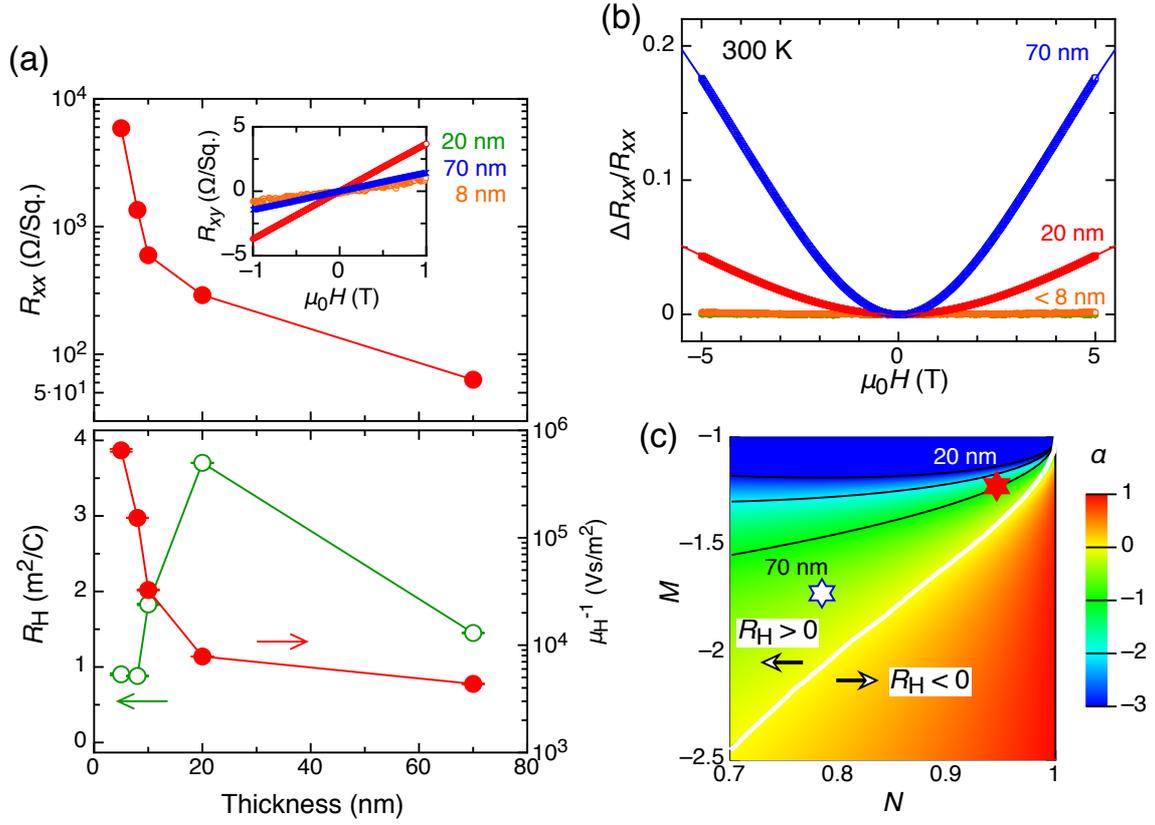

**Figure 2** (a) Thickness dependence of the sheet resistance $R_{xx}$ of Bi/YIG at 300 K. The Hall resistance $R_{xy}$ is shown in the inset. The thickness dependences of the sheet Hall coefficient $R_H = R_{xy}/\mu_0 H$ and the inverse Hall mobility $\mu_H^{-1} = (R_H/R_{xx})^{-1}$ are also presented. (b) Magnetoresistance $[R_{xx}(H) - R_{xx}(0)]/R_{xx}(H)$ of the Bi/YIG at 300 K. The fitting results using the two-carrier model are also displayed as a solid line. (c) Contour plot of Hall factor $\alpha = R_H/R_0$ determined from the two-carrier model, and the value of $\alpha$ determined from the magnetoresistance. The results indicate a dominance of the electron conductions. See the text for details.

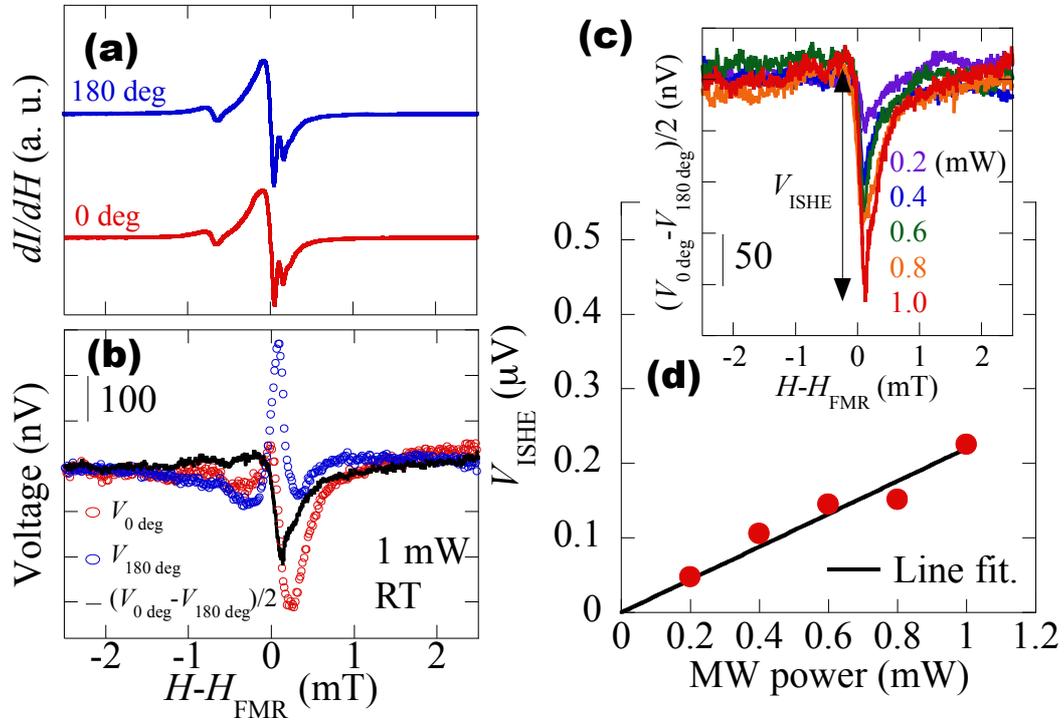

**Figure 3** (a) FMR spectra of the YIG when the external magnetic field is set to be 0 (a red solid line) and 180 (a blue solid line) degrees, respectively. The microwave power was 1 mW. (b) Electromotive forces from the Bi/YIG under the FMR of the YIG. Red and blue open circles are experimental data at the magnetic field of 0 and 180 degrees, respectively. To subtract the thermal effect, the data are averaged (shown as a black solid line). (c) and (d) Excitation power dependence of the electromotive forces from the Bi/YIG. The data were well fitted to a linear function.

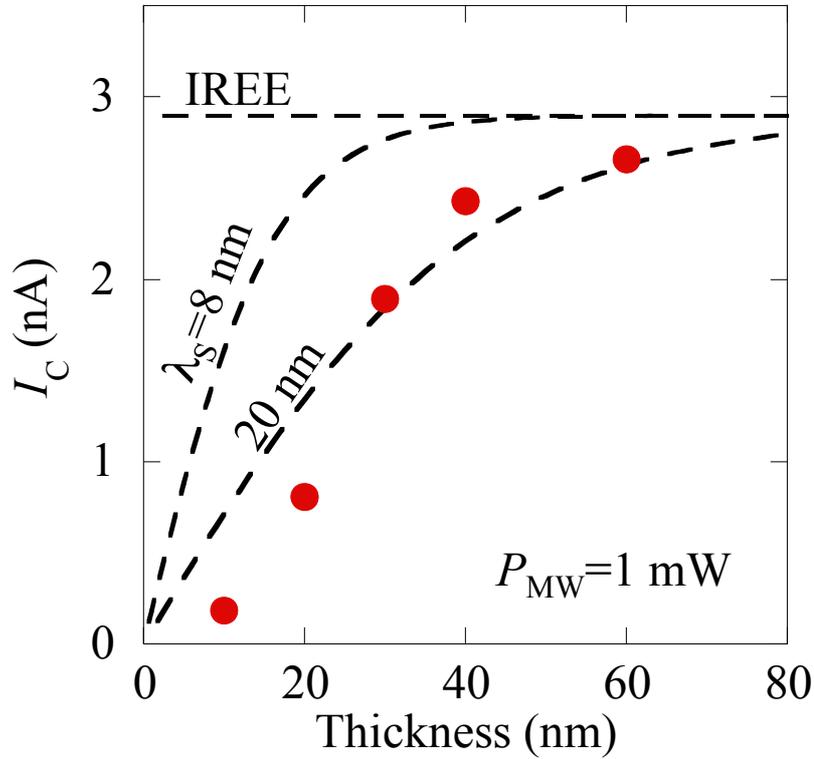

**Figure 4** Thickness dependence of the generated charge current in the Bi/YIG. A monotonic increase in the charged current is clearly observed, indicating the lack of the IREE. The dashed lines labeled "IREE", "$\lambda_s$=8 nm" and "20 nm" are the theoretical lines when we assume that the generation of the charge current is due to the IREE, the ISHE (the spin diffusion length of the Bi is 8 nm), and the ISHE (its spin diffusion length is 20 nm), respectively. The line "IREE" is a line assuming a certain IREE length scale.